\documentclass[aps, prl, showpacs,
superscriptaddress , unsortedaddress, twocolumn,
preprintnumbers]{revtex4}
\usepackage{pstricks, pst-plot}
\usepackage{graphicx}
\newcommand{\Laguerre}{\mathop{\mathrm{L}}\nolimits}
\newcommand{\Tr}{\mathop{\mathrm{Tr}}\nolimits}

\begin{document}
\preprint{PHYSICAL REVIEW A {\bf 80}, 013813 (2009)}
\title{Numerical reconstruction of photon-number statistics from
photocounting statistics:\\  Regularization of an ill-posed problem}

\author{V. N. Starkov}
\affiliation{Institute of Physics, National Academy of Sciences of
Ukraine, Prospect Nauky 46, UA-03028 Kiev, Ukraine}%
\author{A. A. Semenov}
\email[E-mail address: ]{sem@iop.kiev.ua}%
\affiliation{Institute of
Physics, National Academy of Sciences of Ukraine, Prospect Nauky 46,
UA-03028 Kiev, Ukraine}%
\affiliation{Institute of Physics and Technology, National Technical
University of Ukraine ``KPI,''
Prospect Peremohy 37, UA-03056 Kiev, Ukraine}%
\affiliation{Bogolyubov Institute for Theoretical Physics, National
Academy of Sciences of Ukraine,\\ Vul. Metrologichna 14-b, UA-03680
Kiev, Ukraine}

\author{H. V. Gomonay}
\affiliation{Institute of Physics and Technology, National Technical
University of Ukraine ``KPI'', Prospect Peremohy 37, UA-03056 Kiev,
Ukraine}%
\affiliation{Bogolyubov Institute for Theoretical Physics, National
Academy of Sciences of Ukraine,\\ Vul. Metrologichna 14-b,
UA-03680 Kiev, Ukraine}%

\begin{abstract}
We demonstrate a practical possibility of loss compensation in
measured photocounting statistics in the presence of dark counts and
background radiation noise. It is shown that satisfactory results
are obtained even in the case of low detection efficiency and large
experimental errors.
\end{abstract}


\maketitle

Photoelectric detection of quantum light~\cite{PhotoDetection,
MandelBook} is a basic experimental technique in a variety of
fundamental and applied investigations. In principle, the
photon-number resolved detectors enable one to determine the number
of photons in radiation fields. In practice, the number of
photocounts may significantly differ from the number of photons due
to losses, dark counts, and background radiation. The modern
technologies enable one to get the detection efficiency near 0.9 and
even more \cite{Waks}. However, such an improvement leads, as rule,
to an increase in the dark count rate \cite{Takeuchi}. Furthermore,
different losses occur at all the stages of generation,
manipulation, and transmission of quantum light.

The effects of losses and noise in photocounting statistics can be
compensated in two different ways. First, the active compensation
can be realized by using the homodyne detection and preamplification
of the signal by a degenerate parametric amplifier~\cite{Leonhardt}.
Another possibility is a numerical manipulation with the measured
data -- the corresponding technique of loss compensation has been
discussed in Ref.~\cite{Kiss}. The problem is that the corresponding
series can diverge for small efficiency $\eta$ in many important
cases that require application of a special technique of the
analytical continuation~\cite{Herzog}.

However, the most serious problem is that the method of numerical
compensation occurs to be unstable with respect to small
experimental inaccuracies. Small experimental errors in
photocounting statistics may lead to large errors in the
reconstructed photon-number statistics even for large detection
efficiencies. This problem should be resolved by application of
special regularization methods. For example, the photon-number
statistics of a laser radiation has been reconstructed by the method
of maximum entropy in Ref.~\cite{Lee}. The least-squares
regularization technique for loss compensation in photocounting
statistics has been recently considered in the context of the
tomography of the quantum detectors \cite{Lundeen}. The numerical
compensation of losses in multi-pixel detectors has been discussed
in \cite{Afek}. The method of maximum-likelihood estimation
demonstrates satisfactory results for quantum-state reconstruction
in the presence of losses \cite{Mogilevtsev} and, consequently, it
can be applied for loss compensation in photocounting statistics. An
alternative technique of the regularization, which requires
measurements with different values of the efficiency, has been
proposed in Ref.~\cite{Zambra}.

In the present contribution we reexamine the method of numerical
compensation of losses. We  demonstrate that the regularization of
the corresponding ill-posed problem (see, e.g.,~\cite{Tikhonov} and
\cite{Morozov}) enables one to apply this technique even for low
values of the efficiency $\eta$. Moreover, our consideration
includes numerical compensation of dark counts and effects of
background radiation. Besides, the method demonstrates satisfactory
results under a realistic assumption that the efficiency and the
noise-counts rate are known with a certain inaccuracy. We
demonstrate that the proposed technique can be applied for different
photocounting statistics including highly-nonclassical cases.

Let us start with consideration of a single-mode quantum light
characterized by the density operator $\hat\varrho$. If $\hat n$ is
the corresponding photon-number operator and $\left|n\right\rangle$
is its eigenvector, the photon-number distribution is given
by~\cite{PhotoDetection, MandelBook}
\begin{equation}
p_n=\left\langle n \left|
\hat{\varrho}\right| n \right\rangle=\Tr\left(:\frac{\hat{n}^n}{n!}\, \exp\left(-\hat{n}\right):\,
\hat{\varrho}\right).\label{photon-number}
\end{equation}
In the presence of losses, dark counts, and background radiation it
differs from the photocounting distribution~\cite{Karp, Semenov},
\begin{equation}
\mathcal{P}_n=\Tr\left(:\frac{\left(\eta\,\hat{n}\,+N_\mathrm{nc}\right)^n}{n!}\,
\exp\left(-\eta\,\hat{n}-N_\mathrm{nc}\right):\,
\hat{\varrho}\right),\label{photocounts}
\end{equation}
where $\eta$ and $N_\mathrm{nc}$ are the efficiency and the mean
number of noise counts, respectively. The aim of this work is to
develop a mathematical technique for reconstruction of the
photon-number distribution $p_n$ from the experimentally-measured
photocounting distribution $\mathcal{P}_n$.

The photocounting distribution $\mathcal{P}_n$ is expressed in terms
of the photon-number distribution $p_n$ as (see~\cite{Semenov, Lee})
\begin{equation}
\mathcal{P}_m=\sum\limits_{n=0}^{+\infty}S_{m|n}\left(\eta,
N_\mathrm{nc}\right)\, p_n.\label{ProbabilitySeries}
\end{equation}
Here,
\begin{equation}
S_{m|n}\left(\eta, N_\mathrm{nc}\right)= e^{-N_\mathrm{nc}}
N_\mathrm{nc}^{m-n}\eta^{n}\frac{n!}{m!}\Laguerre_n^{m-n}\!
\left(\frac{N_\mathrm{nc}(\eta-1)}{\eta}\right)\label{CondProbPois1}
\end{equation}
for $m\geq n$ and
\begin{equation}
S_{m|n}\left(\eta, N_\mathrm{nc}\right)=
e^{-N_\mathrm{nc}}(1-\eta)^{n-m}\eta^{m}\Laguerre_m^{n-m}\!
\left(\frac{N_\mathrm{nc}(\eta-1)}{\eta}\right)\label{CondProbPois2}
\end{equation}
for $m\leq n$ are the probabilities to get $m$ photocounts under the
condition that $n$ photons are present. $\Laguerre_n^{m}\!
\left(x\right)$ is the Laguerre polynomial.
Expression~(\ref{ProbabilitySeries}) is formally inverted as
\begin{equation}
p_n=\sum\limits_{m=0}^{+\infty}S_{n|m}^{-1}\left(\eta,
N_\mathrm{nc}\right)\, \mathcal{P}_m,\label{ProbabilityInvSeries}
\end{equation}
where \setlength\arraycolsep{0pt}
\begin{eqnarray}
&&S_{n|m}^{-1}\left(\eta,
N_\mathrm{nc}\right)=\frac{1}{\eta^{n}}\Phi
\left(n+1,n-m+1;\frac{N_\mathrm{nc}(1-\eta)}{\eta}\right)\nonumber\\
&&\times
e^{N_\mathrm{nc}}\frac{(-N_\mathrm{nc})^{n-m}}{(n-m)!}\label{CondProbInv1}
\end{eqnarray}
for $m\leq n$ and
\begin{eqnarray}
&&S_{n|m}^{-1}\left(\eta, N_\mathrm{nc}\right)=e^{N_\mathrm{nc}}
\Phi
\left(m+1,m-n+1;\frac{N_\mathrm{nc}(1-\eta)}{\eta}\right)\nonumber\\
&&\times{m\choose
n}\frac{1}{\eta^{n}}\left(1-\frac{1}{\eta}\right)^{m-n}\label{CondProbInv2}
\end{eqnarray}
for $m\geq n$ is the matrix inverse to $S_{m|n}\left(\eta,
N_\mathrm{nc}\right)$. $\Phi\left(n,m;x\right)$ is the Kummer
hypergeometric function.

As mentioned, expression~(\ref{ProbabilityInvSeries}) as the
solution of Eq.~(\ref{ProbabilitySeries}) is unstable with respect
to small experimental inaccuracies of $P_n$. Moreover, similar to
the case of zero noise counts~\cite{Kiss, Herzog}, this series
diverges in many important cases. Hence,
Eqs.~(\ref{ProbabilityInvSeries})--(\ref{CondProbInv2}) cannot be
applied in the general case. This fact is a consequence of a more
general statement that Eq.~(\ref{ProbabilitySeries}) is an ill-posed
problem~\cite{Tikhonov}. Such a problem can be treated by using the
appropriated regularization methods~\cite{Tikhonov, Morozov, Engl}.

{\em A priori} information, such as numbers at which photon-number
and photocounting distributions can be truncated~\cite{Welsch}, is
used in the regularization of the ill-posed problem. In addition,
the basic properties of the photon-number distribution,
\begin{eqnarray}
p_n\geq 0,\label{Positivity}\\
\sum\limits_{n}p_n=1,\label{Norma}
\end{eqnarray}
are applied in the considered case. Other {\em a priori} information
can also be useful for the regularization of the ill-posed problem
depending on a given physical situation.

In different problems of quantum optics the least-squares inversion
and the Tikhonov regularization lead to satisfactory results (for a
review see, e.g., \cite{Welsch}). In this contribution, we apply the
Landweber algorithm \cite{Landweber} adopted to the regularization
of similar problems \cite{Engl} -- a technique, which demonstrates a
good computer compatibility \cite{Brodyn}. The projected Landweber
algorithm~\cite{Byrne} is the iteration process,
\begin{equation}
p^{(j)}=\Pi_C\left[
p^{(j-1)}+\chi\left(S^{\dagger}\,\mathcal{P}-
S^{\dagger}S\,p^{(j-1)} \right) \right]
,\label{Landweber}
\end{equation}
where $p^{(j)}=\left\{p_n^{(j)}\right\}$ is the $j\textrm{th}$
iteration for the photon-number distribution,
$\mathcal{P}=\left\{\mathcal{P}_m\right\}$,
$S=\left\{S_{m|n}\right\}$, and $\chi$ is the relaxation parameter.
$\Pi_C$ is the projector on the closed convex set $C$ defined by
Eq.~(\ref{Positivity}) and, in special cases, by other additional
conditions. Condition~(\ref{Norma}) can be used to track the
accuracy of the obtained results. The starting values are usually
chosen as $p_n^{(0)}=0$.

\begin{figure}[ht!]
\includegraphics[clip=,width=0.95\linewidth]{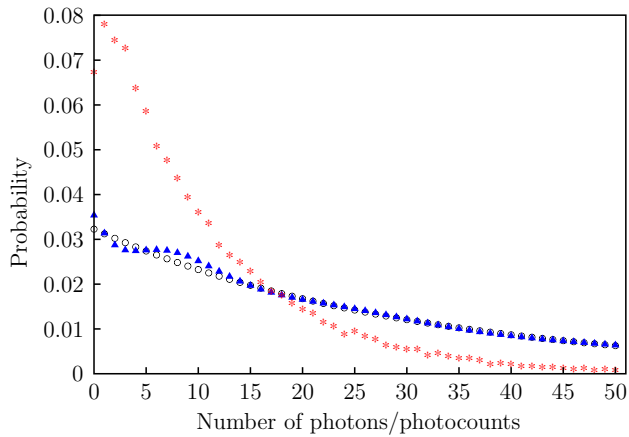}
\caption{\label{Fig4} (Color online) The photon-number, $p_n$, and
photocounting, $\mathcal{P}_n$, distributions of the thermal state,
$\bar{n}_\mathrm{th}=30$. The circles and triangles are the
initially chosen and reconstructed ($\tilde\eta=0.35$,
$\tilde{N}_\mathrm{nc}=0.29$) photon-number distributions,
respectively. The asterisks show the simulated photocounting
distribution ($\nu=5\times 10^4$, $\eta=0.34$,
$N_\mathrm{nc}=0.30$).}
\end{figure}

To illustrate the method let us give some numerical simulations. We
start from the thermal state,
\begin{equation}
\hat\varrho=\frac{1}{1+\bar{n}_\mathrm{th}}
\left(\frac{\bar{n}_\mathrm{th}}{1+\bar{n}_\mathrm{th}}\right)^{\hat{n}}\label{TS_DO}
\end{equation}
with $\bar{n}_\mathrm{th}=30$, and derive from
Eq.~(\ref{photocounts}) the photocounting distribution
$\mathcal{P}_n$ for $\eta=0.34$ and $N_\mathrm{nc}=0.30$. The
measured data are simulated with $\nu=5\times 10^4$ sampling events.
The corresponding relative error (in terms of the Euclidian norm) is
$\delta_P=0.03$. The simulated data are then used as an input of
Landweber algorithm~(\ref{Landweber}) for the reconstruction of the
photon-number distribution with inaccurate values of the efficiency
$\tilde\eta=0.35$ and the mean number of noise counts
$\tilde{N}_\mathrm{nc}=0.29$. The result of this procedure (see
Fig.~\ref{Fig4}) is in a reasonable agreement with the initially
chosen photon-number distribution, the relative error is
$\delta_p=0.05$, and the relative residual is
$\tilde{\delta}_p=0.019$. It is worth noting that in the given
example series~(\ref{ProbabilitySeries}) diverges even in the
absence of experimental errors~\cite{Kiss}. Nevertheless,
algorithm~(\ref{Landweber}) demonstrates high efficiency for the
considered case.

\begin{figure}[ht!]
\includegraphics[clip=,width=0.95\linewidth]{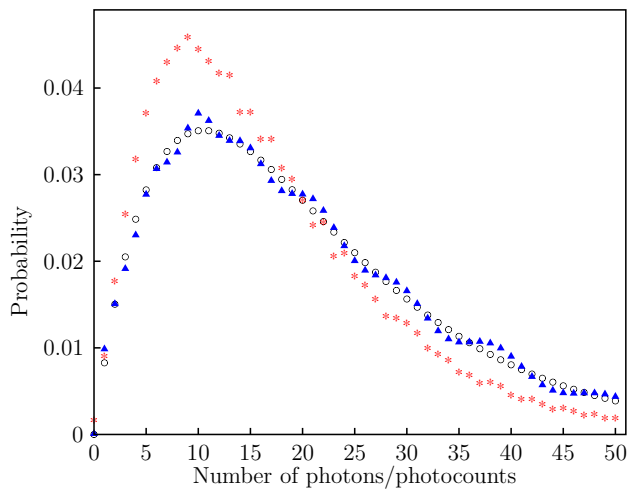}
\caption{\label{Fig2} (Color online) The photon-number, $p_n$, and
photocounting, $\mathcal{P}_n$, distributions of the SPATS,
$\bar{n}_\mathrm{th}=10$. The circles and triangles are the
initially chosen and reconstructed ($\tilde\eta=0.77$,
$\tilde{N}_\mathrm{nc}=0.75$) photon-number distributions,
respectively. The asterisks show the simulated photocounting
distribution ($\nu=5\times 10^3$, $\eta=0.7764$,
$N_\mathrm{nc}=0.748$).}
\end{figure}

\begin{figure}[ht!]
 \includegraphics[clip=,width=0.95\linewidth]{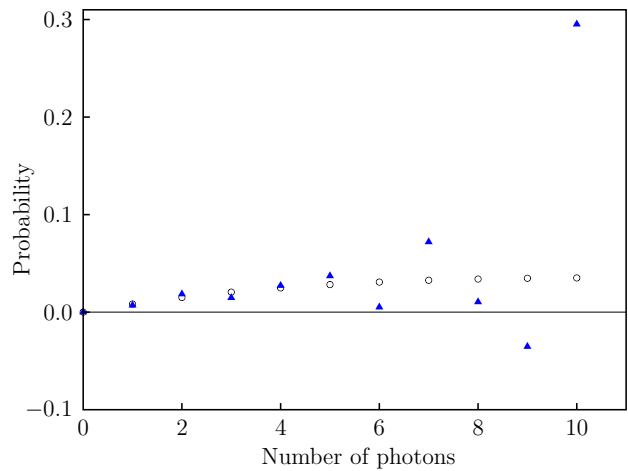}
 \caption{\label{Fig1} (Color online) The photon-number distribution, $p_n$, of the
 SPATS, $\bar{n}_\mathrm{th}=10$. The circles and triangles are the
initially chosen and reconstructed ($\nu=5\times 10^5$,
$\eta=0.7764$, ${N}_\mathrm{nc}=0.748$) distributions in
Eq.~(\ref{ProbabilityInvSeries}), respectively. The result of such a
reconstruction demonstrates much stronger noise effect in comparison
with the Landweber algorithm (see Fig.~\ref{Fig2}).}
 \end{figure}

Another example is the single-photon-added thermal state (SPATS),
 \begin{equation}
 \hat\varrho=\frac{\hat{n}}{\bar{n}_\mathrm{th}\left(1+\bar{n}_\mathrm{th}\right)}
 \left(\frac{\bar{n}_\mathrm{th}}{1+\bar{n}_\mathrm{th}}\right)^{\hat{n}}\label{SPATS_DO}
 \end{equation}
with $\bar{n}_\mathrm{th}=10$. Such a state has been recently
realized experimentally and its nonclassical properties have been
verified~\cite{SPATS}. The numerical simulation is performed for
$\nu=5\times 10^3$, $\eta=0.7764$, and $N_\mathrm{nc}=0.748$, with
the relative error $\delta_P=0.025$. The photon-number distribution
(reconstructed with $\tilde\eta=0.77$, $\tilde{N}_\mathrm{nc}=0.75$)
is shown in Fig.~\ref{Fig2}. The relative error is $\delta_p=0.041$
and the relative residual is $\tilde{\delta}_p=0.020$. In this
example series~(\ref{ProbabilityInvSeries}) formally converges.
However, attempts to apply this series for the direct reconstruction
of $p_n$ result in the large noise effects caused by small
experimental inaccuracies even for sufficiently large number of
sampling events ($\nu=5\times 10^5$, $\delta_P=0.0079$) and exact
values of $\eta$ and $N_\mathrm{nc}$ (see Fig.~\ref{Fig1}).

\begin{figure}[ht!]
\includegraphics[clip=,width=0.95\linewidth]{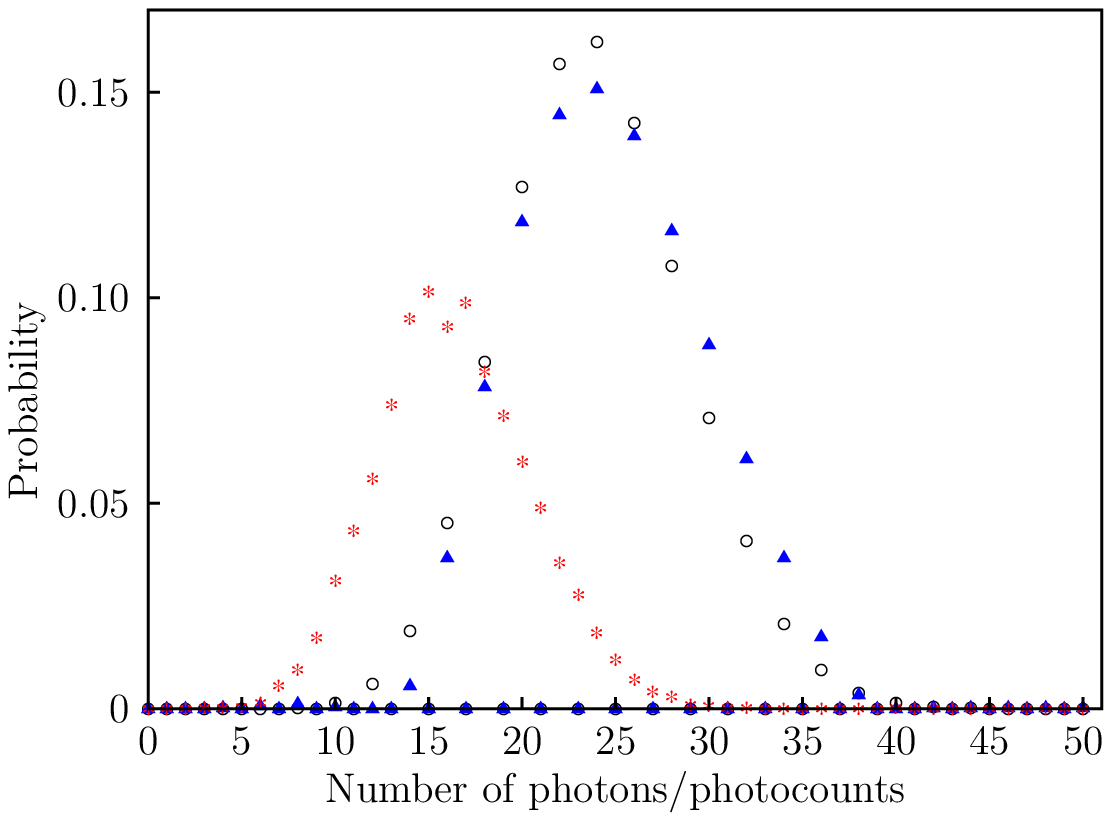}
\caption{\label{Fig3} (Color online) The photon-number, $p_n$, and
photocounting, $\mathcal{P}_n$, distributions of the superposition
of the coherent states $\left|\alpha\right|^2=23.9$. The circles and
triangles are the initially chosen and reconstructed (
 $\tilde\eta=0.59$,
$\tilde{N}_\mathrm{nc}=1.77$) photon-number distributions,
respectively. The asterisks show the simulated photocounting
distribution ($\nu=5\times 10^3$, $\eta=0.613749$,
$N_\mathrm{nc}=1.763442$).}
\end{figure}

Algorithm~(\ref{Landweber}) demonstrates a reasonable agreement with
the initially chosen photocounting distribution even in the case of
rather large error in the simulated data. To illustrate this fact,
we consider the superposition of the coherent states,
\begin{equation}
\left|\psi\right\rangle=\frac{1}{\sqrt{2\left(1+e^{-2\left|\alpha\right|^2}\right)}}
\Big(\left|\alpha\right\rangle+\left|-\alpha\right\rangle\Big),
\label{OSCS}
\end{equation}
$\left|\alpha\right|^2=23.9$. The initial data are simulated with
$\nu=5\times 10^3$, $\eta=0.613749$, and $N_\mathrm{nc}=1.763442$.
The relative error is $\delta_P=0.098$. The photon-number
distribution is reconstructed for $\tilde\eta=0.59$,
$\tilde{N}_\mathrm{nc}=1.77$ (see Fig.~\ref{Fig3}). The relative
error of the reconstructed distribution, $\delta_p=0.125$, and the
relative residual, $\tilde{\delta}_p=0.051$, are of the same order
as for the initially simulated data. It  should be stressed that for
the state~(\ref{OSCS}) $p_n\neq 0$ only for even photon numbers $n$.
This {\em a priori} information is used in the projector $\Pi_C$,
[cf.~Eq.~(\ref{Landweber})].

In conclusion, we have obtained the expression for photocounting
distribution in terms of the photon-number distribution. However,
the inverted expression cannot be used in the most practical
situations -- it is unstable with respect to small experimental
inaccuracies and the corresponding series can diverge. At the same
time, the regularization of this ill-posed problem by the Landweber
algorithm enables one to compensate losses and noise counts even for
low efficiencies, high noise-counts rates, and inaccurate knowledge
of their values.

\acknowledgements

The authors acknowledge support by the Fundamental Researches State
Fund of Ukraine. A.A.S. also thanks NATO Science for Peace and
Security Programme for financial support. The authors thank S.L.
Braunstein for providing with references regarding an alternative
regularization method of the considered problem.

\end{document}